# Learned optical compression for single-pixel quantitative phase imaging

Ary Portes[1,2,3†], Yuntian Wang[1,2,3†], Che-Yung Shen[1,2,3], Shiqi Chen[1,2,3] and Aydogan Ozcan[1,2,3*]

[1]Electrical and Computer Engineering Department, University of California, Los Angeles, CA, 90095, USA
[2]Bioengineering Department, University of California, Los Angeles, CA, 90095, USA
[3]California NanoSystems Institute (CNSI), University of California, Los Angeles, CA, 90095, USA

† These authors contributed equally to this work
* Corresponding author: ozcan@ucla.edu

# Abstract

Quantitative phase imaging (QPI) reveals optical path-length variations in transparent objects but typically relies on spatially resolved intensity measurements. Here we numerically and experimentally demonstrate learned optical compression for single-pixel QPI, jointly optimizing a diffractive encoder and a digital neural decoder to concentrate reconstruction-relevant phase information into a compact measurement vector. A single reconfigurable phase-modulation plane sequentially displays the learned encoding patterns, which are independent of the input object, each yielding one scalar intensity measurement. Operating at a visible wavelength, the experimental system reconstructs quantitative phase images of unknown objects with ~40-fold dimensional compression, using only 2.55% as many intensity measurements as reconstructed pixels. Experimental blind testing on handwritten digits and grating-like phase objects achieves mean Pearson correlation coefficients of 0.945 and 0.848, respectively. Benchmarking under the same measurement budget demonstrates substantially higher reconstruction fidelity than other encoding strategies, highlighting the advantages of learned optical compression. With the optical encoder fixed, decoder-only neural adaptation enables experimental transfer to previously unseen data distributions, extending the utility of the learned measurement representation. These results establish learned optical compression as a measurement-efficient and reusable framework for QPI, integrating phase-information compression into single-pixel intensity measurements while retaining adaptability to new data. This approach paves the way for low-data-rate phase imaging and sensing, with the potential to extend QPI to spectral regions where high-quality detector arrays are costly or unavailable.

# Introduction

Quantitative phase imaging (QPI) reveals optical path-length variations in transparent specimens, providing label-free contrast associated with refractive index and thickness variations[1–3]. Because detectors record intensity rather than phase, conventional QPI methods convert phase information into intensity variations and computationally invert this mapping, which is generally ill-posed without additional information[1,4,5]. Established approaches obtain this information through, e.g., off-axis holography[6], phase-shifting interferometry[7], common-path interferometry[8,9], or measurement diversity based on defocus[10–12], scanning of illumination angle[13–16], and asymmetric illumination[17]. Deep-learning-based reconstruction methods[18–20] can reduce the required image count[19,21–23], but generally retain dense spatial sampling despite the compressibility of many natural and biological phase objects[24]. Reliance on detector arrays imposes data acquisition, storage and readout-related burdens and limits the use of QPI at wavelengths where suitable detector arrays are costly or unavailable[25,26]. This motivates an acquisition strategy that compresses phase information optically into a small set of scalar measurements, rather than first recording densely sampled intensity images for subsequent digital reconstruction.

Single-pixel imaging avoids spatially resolved detection[25,26] by encoding an object's spatial information in time[26–28] or wavelength[29,30] and reconstructing the target image from scalar measurements[31]. This single-pixel architecture has enabled infrared and terahertz imaging[32,33], but its efficiency depends on which encodings minimize the required measurement count[34]. For intensity imaging, Hadamard and Fourier patterns[27,28] and random compressive measurements[24,26] exploit a linear measurement model. Phase-to-intensity[35] conversion is instead nonlinear, so encodings that are effective for intensity imaging need not efficiently preserve phase information. Existing single-pixel QPI methods use interferometric spatial-frequency projections[36], random phase modulation[37], or Hadamard and Fourier phase patterns[38], prescribed independently of object statistics and digital reconstruction methods. The challenge is to design optical encodings that exploit both the nonlinear phase-to-intensity mapping and object phase statistics to support high-fidelity quantitative phase reconstruction from a small number of scalar, single-pixel measurements.

Diffractive optical surfaces provide a physical route to learned optical encoding, with phase profiles optimized through differentiable wave propagation[39]. For example, diffractive QPI has converted object phase into spatially resolved intensity images[40], including at multiple wavelengths[41] and through unknown/random diffusers[42]. Despite visible-wavelength demonstrations for other optical-processing tasks[43,44], diffractive QPI remains limited in this regime[45], with cascaded surfaces introducing fabrication and alignment demands[46]. Hybrid optical-digital architectures instead compress optical inputs into representations decoded by neural networks[47–50], including spectral measurements acquired by a single-pixel detector[30]. An optical compressor coupled to a digital decoder has also been numerically studied for QPI, without time-multiplexed encoding or experimental implementation[51]. The central opportunity is therefore to learn a compact, experimentally realizable optical measurement representation that preserves quantitative phase information for single-pixel-based QPI reconstructions.

Here, we report learned optical compression for single-pixel QPI, in which the optical acquisition itself is trained to concentrate reconstruction-relevant phase information into a compact set of scalar intensity measurements. By jointly optimizing a diffractive encoder and a digital neural decoder through a differentiable wave-propagation model, we learn measurement patterns tailored to both the object statistics and the physics of phase-to-intensity conversion. A single reconfigurable phase-modulation plane sequentially displays the learned patterns, which are independent of the input phase object to be reconstructed, with optical propagation and intensity detection yielding one scalar measurement per pattern. In blind-testing experiments at 635 nm, we reconstruct 28 × 28-pixel quantitative phase images of previously unseen handwritten digits and grating-like phase objects from only 20 scalar intensity measurements. This corresponds to a measurement-to-output-pixel ratio of 2.55% and ~40-fold dimensional compression. At the same 20-measurement budget, numerical benchmarking with matched decoder architectures and training protocols yields an 11.4-12.7 dB improvement in mean PSNR (peak signal-to-noise ratio) over the tested random, Hadamard and Fourier encoders. This advantage demonstrates the value of jointly learning optical encoding, single-pixel acquisition and digital reconstruction rather than optimizing reconstruction alone for prescribed phase encodings. Beyond measurement efficiency, the same 20 encoder phase patterns also support experimental reconstruction of phase objects drawn from new data distributions that were excluded from their original optimization. Decoder-only retraining reduces the simulation-to-experiment mismatch and enables this transfer without re-optimizing the optical encoder or increasing the per-object measurement count.

Together, these results establish learned optical compression as a measurement-efficient and reusable framework for single-pixel QPI, combining a compact phase measurement representation with digitally adaptable phase reconstruction. By encoding QPI into a compact stream of scalar measurements, this learned optical compression approach opens a path toward low-data-rate phase sensing in biomedical imaging and optical inspection and could extend label-free QPI to parts of the electromagnetic spectrum where high-quality detector arrays are costly or unavailable.

# Results

### Single-pixel QPI using sequential optimized phase encoding

Learned optical compression maps each input phase distribution into a compact sequence of optimized scalar intensity measurements, which a jointly trained neural decoder converts into a quantitative phase image (**Fig. 1**). In the demonstrated configuration, 20 scalar measurements support a 28 × 28-pixel reconstruction, corresponding to a measurement-to-output-pixel ratio of 2.55% and ~40-fold dimensional compression. **Figure 1** summarizes the operating principle of the single-pixel QPI system. As shown in **Fig. 1(a)**, our input phase object (the target sample) is defined as:

$$\Phi_{\text{target}}(x,y) = \alpha\pi I(x,y), \qquad I(x,y) \in [0,1] \tag{1}$$

where $I(x,y)$ is a normalized spatial data distribution and α is a dimensionless scaling factor that sets the maximum target phase to $\alpha\pi$. The target phase object to be reconstructed, $\Phi_{\text{target}}$, remains fixed while it is sequentially superimposed with $M$ optimized phase-only encoder patterns, $\mathcal{E}(x,y,t_i)$, under coherent illumination at wavelength λ, where $i = 1, \dots, M$. The encoder patterns are learned and fixed (i.e., they are *independent* of the unknown object to be reconstructed), and span a phase range of 0–2π. Stated differently, after their optimization the encoder patterns are held fixed for all test objects; no object-specific optical re-optimization is performed. At the single-pixel acquisition at $t_i$, the encoded phase-only optical field, composed of $\exp j[\mathcal{E}(x,y,t_i) + \Phi_{\text{target}}(x,y)]$, propagates to the detector plane, where its intensity is averaged over a detector aperture of width $w_d$ to produce one scalar measurement, $m_i$. Repeating this process for all $i = 1, \dots, M$ encoder states forms the nonnegative measurement vector:

$$\boldsymbol{m} = [m_1, m_2, \dots, m_M]^{\mathrm{T}} \tag{2}$$

As illustrated in **Fig. 1(b)**, the digital neural decoder, $\mathcal{D}(\cdot)$, first normalizes the *M*-element measurement vector and expands it through a sequence of fully connected layers. The resulting features are reshaped into a spatial latent representation and upsampled through two transposed-convolutional stages to reconstruct the phase distribution of the input phase object:

$$\Phi_{\mathrm{r}}(x, y) = \mathcal{D}(\boldsymbol{m}) \approx \Phi_{\mathrm{target}}(x, y) \tag{3}$$

The layer dimensions and intermediate feature-map sizes are depicted in **Fig. 1(b)**, with complete architectural details provided in the **Methods**.

The object-independent encoder phase patterns, $\mathcal{E}(x, y, t_i)$, and the neural decoder, $\mathcal{D}(\cdot)$, were jointly optimized using supervised training within a differentiable model of the complete data acquisition process (see **Fig. 2(a)**). Training phase objects were randomly drawn from the MNIST dataset and a custom-built grating-like image dataset (see the **Methods** section). The phase-range scaling factor was chosen as $\alpha = 1$ to limit the target objects' phase range between 0 and $\pi$ unless otherwise mentioned. The resulting complex field, $\exp j[\mathcal{E}(x, y, t_i) + \Phi_{\mathrm{target}}(x, y)]$, was then propagated through the differentiable optical model onto the detector plane. Spatial averaging of the detector-plane intensity over the finite detector aperture yielded the measurement vector $\boldsymbol{m}$, which was processed by the digital decoder $\mathcal{D}$ to reconstruct the target phase distribution for each training sample. Since the complete optical data acquisition and image reconstruction model was differentiable, gradients were backpropagated through both the optical encoder and the digital decoder, enabling end-to-end joint optimization. The training loss function combined pixelwise mean square error of the phase reconstruction with a detector-plane energy penalty term that promoted optical energy to remain concentrated within the finite detector aperture (see **Methods** for details). The resulting optimized encoder phase patterns, $\mathcal{E}(x, y, t_i)$, $i = 1, \dots, 20$ are shown in **Fig. 2(b)**; post-training, these encoder patterns remain frozen, independent of the input target phase function to be reconstructed, $\Phi_{\mathrm{target}}(x, y)$.

**Experimental single-pixel phase reconstruction at ~40-fold dimensional compression**

The optimized encoder was tested experimentally using the visible-wavelength optical system shown in **Fig. 3**. As illustrated in **Fig. 3(a)**, a 635-nm coherent beam illuminated a reflective phase-only spatial light modulator (SLM), which displayed the phase object together with each of the 20 encoder patterns $\mathcal{E}(x, y, t_i)$ in a sequence. **Fig. 3(b)** shows the corresponding experimental setup. At the detector plane, each encoder pattern produced a spatial intensity distribution, as shown in **Fig. 3(c)**. A fixed region of interest, corresponding to approximately $0.1 \times 0.1\ \mathrm{mm}^2$, defined the single-pixel detection aperture. The optical intensity within this region generated one scalar measurement for each encoder pattern, and repeating this process for all $M = 20$ encoder patterns created the measurement vector $\boldsymbol{m}$ supplied to the digital decoder, i.e., $\mathcal{D}(\boldsymbol{m})$; no spatially resolved camera image was supplied for reconstruction, and the remaining camera pixels were used only for alignment.

For the experimental deployment and testing of the trained model, the $M = 20$ encoder states, $\mathcal{E}_0(x, y, t_i)$, were quantized into 8-bit SLM patterns and fixed for experimental deployment. Zero-shot QPI was performed by applying the simulation-trained decoder $\mathcal{D}$ directly to the experimental measurement vectors, $\Phi_r(x, y) = \mathcal{D}_0(\mathbf{m}_{\mathbf{exp}})$; in this notation, we use $\mathcal{D} \rightarrow \mathcal{D}_0$ to indicate the model was not exposed to experimental training data. This zero-shot deployment achieved mean PCC (Pearson correlation coefficient) values of 0.848 for MNIST and 0.752 for grating-like test objects that were never seen before (**Fig. 4(c,d)**), demonstrating direct transfer from the digital twin to the physical/experimental system. The remaining simulation-to-experiment gap reflects experimental imperfections, including photon noise, misalignment, and detector-sampling inaccuracies. To reduce this gap, $\mathcal{D}_0$ was fine-tuned using $T$ experimentally tested objects from each dataset, yielding $\mathcal{D}_0 \rightarrow \mathcal{D}_{\text{exp}}$ while $\mathcal{E}_0(x, y, t_i)$ remained fixed; for example, $T = 900$ denotes 900 MNIST and 900 grating objects used during this experimental transfer learning phase. Final blind testing through $\mathcal{D}_{exp}(\boldsymbol{m}_{\text{exp}})$ was applied to separate phase-only objects never used before. **Fig. 3(d)** compares simulated $(\boldsymbol{m})$ and experimental $(\boldsymbol{m}_{\text{exp}})$ measurement vectors for a test object held out from decoder training. Their strong correlation across the 20 encoder states $(r = 0.971)$ shows that the digital twin closely reproduces the experimental measurement response. After transfer learning with $T = 900$ objects per dataset, the experimental phase reconstructions $\Phi_r(x, y) = \mathcal{D}_{exp}(\boldsymbol{m}_{\text{exp}})$ achieved a PCC of 0.979 and a PSNR of 24.9 dB, approaching the simulated reconstruction performance of PCC of 0.989 and PSNR of 27.9 dB (**Fig. 3(e)**).

Furthermore, **Fig. 4(a,b)** compares blind-test reconstructions from simulated measurements using $\mathcal{D}(\boldsymbol{m})$, zero-shot experimental measurements using $\mathcal{D}_0(\boldsymbol{m}_{\mathbf{exp}})$, and transfer-learned experimental measurements using $\mathcal{D}_{exp}(\boldsymbol{m}_{\text{exp}})$, while the optimized encoder $\mathcal{E}_0(x, y, t_i)$ remains fixed. The simulated reconstructions closely reproduce both the localized structures of the MNIST objects and the extended periodic features of the grating-like phase objects, which have never been seen before. Under zero-shot deployment, the principal object structures remain recognizable, showing that $\mathcal{D}_0$ can recover meaningful phase information directly from experimental measurements. However, these reconstructions $\mathcal{D}_0(\boldsymbol{m}_{\mathbf{exp}})$ exhibit relatively attenuated phase contrast, local spatial distortions, and background artifacts, reflecting the mismatch between the digital twin and the experimental set-up. Transfer learning substantially reduces these errors: using $\Phi_r(x, y) = \mathcal{D}_{exp}(\boldsymbol{m}_{\text{exp}})$ the reconstructed object boundaries and grating features more closely follow the ground truth, phase contrast is better recovered, and background artifacts are suppressed. This improvement across both object classes demonstrates that much of the experimental mismatch compared to the digital twin can be corrected through $\mathcal{D}_0 \rightarrow \mathcal{D}_{\text{exp}}$ without modifying $\mathcal{E}_0(x, y, t_i)$.

To statistically quantify the phase reconstruction quality, PCC and PSNR values between the reconstructed and ground-truth phase distributions were reported through the mean ± standard deviation values over the blind test set in **Fig. 4(c)**. For MNIST test objects, the simulated PCC of 0.973 ± 0.025 and PSNR of 25.7 ± 4.6 dB fell to 0.848 ± 0.056 and 16.8 ± 1.3 dB, respectively, when the same decoder was applied directly to experimental measurements, $\mathcal{D}_0(\boldsymbol{m}_{\mathbf{exp}})$. Transfer learning with $T = 100$ experimental objects per dataset recovered most of the gap (PCC 0.939 ± 0.036, PSNR 21.1 ± 2.4 dB), while a further increase of experimental transfer learning to $T = 900$ improved the performance to 0.945 ± 0.034 (PCC) and 22.2 ± 2.9 dB (PSNR). On the other hand, the grating test dataset encompassed a broad range of spatial periods, orientations, and pattern geometries (see **Methods**; representative examples are shown in **Fig. 4(b)**). Its phase reconstruction performance is summarized in **Fig. 4(d)**. The mean simulated reconstruction PCC was 0.956 ± 0.140 (PSNR 34.1 ± 8.8 dB), while zero-shot experimental deployment achieved 0.752 ± 0.260 (15.7 ± 3.4 dB), respectively. Experimental transfer learning further increased the mean PCC to 0.848 ± 0.233 and the mean PSNR to 19.6 ± 4.3 dB at $T = 900$, while all 20 optical encoder patterns remained fixed. Although the experimental reconstructions exhibited greater variability than those of MNIST test objects, decoder adaptation produced a consistent improvement in the single-pixel QPI fidelity.

Together, these blind tests demonstrate ~40-fold dimensional compression for phase objects ranging from sparse, localized handwritten digits to extended periodic gratings, using a single optimized optical encoder, scalar intensity detection and a jointly optimized digital decoder. Reconstruction of held-out objects before experimental fine-tuning establishes direct simulation-to-experiment transfer, while decoder-only adaptation substantially improves fidelity without changing the optical encoding patterns. The same compact measurement representation therefore supports both structurally distinct object classes and experimental adaptation of their reconstruction.

To further shed light on how the single-pixel QPI reconstruction fidelity depends on the spatial feature size of test objects, we evaluated linear grating phase objects with periods between 142.9 and 714.3 μm and orientations spanning 0–180° (**Fig. 5**). Each test grating, never seen before, was reconstructed in simulation and from its experimental measurements, both zero-shot $\mathcal{D}_0(\boldsymbol{m}_{\mathbf{exp}})$ and after transfer learning $\mathcal{D}_{exp}(\boldsymbol{m}_{\text{exp}})$ with $T = 900$. For a fair comparison, all gratings shown in **Fig. 5** were held out from the learning sets. **Fig. 5(a)** presents representative ground-truth, simulated, and experimental reconstructions for grating periods of 657, 571, 429, and 229 μm. **Fig. 5(b)** compares their phase profiles, extracted along a line perpendicular to the grating fringes, together with the corresponding phase errors. The simulated profiles closely follow the ground truth for all four gratings (PCC ≥ 0.996). Both sets of experimental reconstructions recover the grating structure: zero-shot PCC is 0.963–0.966 for the three coarser gratings and 0.899 for the 229 μm grating, rising to 0.985–0.992 and 0.926, respectively, after transfer learning, $\mathcal{D}_{exp}(\boldsymbol{m}_{\text{exp}})$. The phase errors nevertheless grow as the period decreases. This period dependence of single-pixel QPI is quantified in **Fig. 5(c)**, where all curves are evaluated on the same 54 held-out linear gratings grouped into seven period bins. In simulation results $\mathcal{D}(\boldsymbol{m})$, the phase reconstruction fidelity is nearly independent of period: the mean PCC is 0.978 in the shortest-period bin (171 μm) and 0.998–0.999 for periods of 263 μm and longer. The zero-shot experimental reconstructions $\mathcal{D}_0(\boldsymbol{m}_{\text{exp}})$ show a markedly different trend, reaching mean PCC values of 0.832–0.925 at long periods but falling to 0.351 in the shortest-period bin. To test whether measurement-domain perturbations could account for this loss, we approximated an additive Gaussian perturbation model using the residuals between measured and simulated detector signals for the same objects (see **Methods**), obtaining $\sigma_{\text{m}} = 0.313$ relative to the spread of each object's measurement vector. Applying these perturbations to the simulated measurements and decoding them with the unchanged decoder $\mathcal{D}_0$ reproduced much of the experimental period dependence: the mean PCC decreased from 0.924–0.941 at long periods to 0.410 in the shortest-period bin, closely following the zero-shot results, $\mathcal{D}_0(\boldsymbol{m}_{\text{exp}})$. Transfer learning, $\mathcal{D}_{exp}(\boldsymbol{m}_{\text{exp}})$, recovered most of the long-period degradation, increasing the mean PCC to 0.941–0.975 for periods of 263 μm and longer.

At the shortest period, however, the improvement was modest, from 0.351 to 0.447, leaving performance below the noise-free simulation results.

**Learned optical encoding improves single-pixel QPI fidelity under a fixed measurement budget**

Next, we investigated whether the optimized single-pixel measurement process outperforms conventional encoding strategies under the same measurement budget. To assess the benefit of jointly optimizing the optical encoder and digital decoder, we numerically compared the optimized encoder with random, Hadamard, and Fourier phase-pattern baselines (see **Methods**). These baselines represent established encoding strategies in single-pixel imaging and single-pixel QPI, including random projections[26,52], spatial-frequency projections[36], and fixed Hadamard or Fourier phase bases[38]. Unlike the encoder presented in this work, these measurement patterns are prescribed independently of the object data distribution and digital reconstruction model. This comparison therefore tests whether optimizing the physical measurement process enables a limited set of scalar measurements to preserve more information for quantitative phase reconstruction than established fixed-pattern encoding strategies.

The benchmarking pipeline is summarized in **Fig. 6**. Each encoder contained $M = 20$ phase masks and therefore generated the same number of scalar measurements per unknown test object. The optimized condition used the original jointly optimized encoder and decoder, $\mathcal{E}_0$ and $\mathcal{D}_0$. For each fixed-pattern baseline, the random, Hadamard, or Fourier encoder was frozen and paired with a newly initialized digital decoder having the same architecture as $\mathcal{D}_0$. Each baseline decoder was optimized independently under the same training and evaluation protocol. The comparison therefore evaluates the complete jointly optimized optical-digital system against systems employing prescribed optical encoders and separately optimized digital reconstruction. The mask construction and complete training protocol are detailed in the **Methods** section.

Representative blind-testing reconstructions and the corresponding quantitative metrics are shown in **Fig. 7**. The optimized encoder $\mathcal{E}_0$ preserved the structures of both MNIST and grating-like phase objects more faithfully than the other encoder baselines, achieving a mean PCC of 0.969 ± 0.077 and a PSNR of 29.9 ± 7.9 dB. By comparison, the random encoders achieved 0.811 ± 0.233 PCC and 18.5 ± 4.7 dB PSNR, while the Hadamard and Fourier encoders achieved PCC values of 0.804 ± 0.224 and 0.789 ± 0.222 and PSNR values of 18.0 ± 4.1 and 17.2 ± 4.0 dB, respectively. The jointly optimized system improved mean PSNR by 11.4, 11.9 and 12.7 dB over the random, Hadamard and Fourier baselines, respectively, while using the same 20-measurement budget and neural decoder architecture. These results demonstrate that learning the optical measurements together with the decoder provides a substantial reconstruction advantage for single-pixel QPI over optimizing the neural decoder alone for the tested prescribed phase encodings.

Next, we examined how reconstruction performance changes with the number ($M$) of optimized encoder states. This analysis separates the benefit of the encoder design from the effect of simply acquiring additional measurements. The number of sequential scalar measurements, $M$, determines both the optical compression and the reconstruction fidelity of the proposed system. Reducing M shortens the acquisition sequence and increases compression, but also limits the phase information available to the digital decoder. To characterize this trade-off, **Fig. S1** evaluates reconstruction performance for $M \in \{2,3,5,10,20,50\}$ using various test phase objects. PCC and PSNR are reported while the output resolution remains fixed at $N \times N = 28 \times 28$ pixels. Therefore, the observed changes in fidelity arise from the number of acquired scalar measurements rather than from changes in reconstruction resolution. For this analysis, we define the measurement ratio $\rho$ as

$$\rho = \frac{M}{N^2} \tag{4}$$

Across the sweep, the measurement ratio ranges from $2/784$ to $50/784$, corresponding to approximately $0.26\% - 6.38\%$. Both PCC and PSNR values improve markedly at low measurement counts and continue to increase as M approaches 20. Grating-like test objects exhibit the largest reduction in fidelity under lower values of $\rho$, reflecting the greater measurement diversity required to recover their spatially varying phase structure, whereas MNIST test objects remain more recognizable at smaller $\rho$. Increasing $M$ from 20 to 50 provides further reconstruction gains, whereas $M = 20$ offers a practical balance between QPI fidelity and sequential measurement count under the evaluated training protocol. We used this 20-measurement configuration for experimental deployment, corresponding to a 2.55% measurement-to-output-pixel ratio and ~40-fold dimensional compression.

**Generalization and transferability of the frozen optical encoder to new data distributions**

Using numerical simulations, we next investigated whether the frozen, optimized optical encoder could support object phase ranges beyond the $0 - \pi$ range used during its original joint optimization, and whether decoder-only fine-tuning could broaden the operating range of the single-pixel QPI system. As illustrated in **Fig. 8**, in this analysis the original digital decoder $D_0$ is hereafter referred to as the "Phase Specialist" model. This decoder was jointly optimized with the optical encoder only at $\alpha_{\text{train}} = 1$, corresponding to a maximum object phase of $\pi$. Phase Generalist I and Phase Generalist II were initialized from the Phase Specialist and fine-tuned using phase-scale factors sampled from $\alpha \in [0.5,1.5]$ and $\alpha \in [0,2]$, respectively. In both cases, the decoder was fine-tuned with its output range expanded from [0, π] to $[0, 2\pi]$, while the optical encoder $\mathcal{E}_0$ remained *fixed*. The three decoders were subsequently evaluated using the same previously unseen test objects over $\alpha_{\text{test}} \in [0.125,2]$.

Representative MNIST test object reconstructions are shown in **Fig. 9(a)**, while **Fig. 9(b)** reports the mean PCC over MNIST test objects. Despite being optimized only at $\alpha = 1$, the Phase Specialist retained high structural fidelity outside its optimization condition, achieving mean PCC values from 0.913 to 0.974 over $\alpha_{\text{test}} = 0.625 - 1.375$. Phase Generalist I extended this range further, maintaining a mean PCC above 0.910 over $\alpha_{\text{test}} = 0.375 - 1.625$. In particular, it achieved PCC values of 0.934 at $\alpha_{\text{test}} = 0.375$ and 0.914 at $\alpha_{\text{test}} = 1.625$, both outside its fine-tuning range. These results demonstrate external phase-range generalization for Phase Generalist I, which retained high structural correlation at phase scales outside its fine-tuning interval. Phase Generalist II, whose fine-tuning range encompassed the complete evaluation interval, achieved a PCC above 0.910 from $\alpha_{\text{test}} = 0.25$ to 2, including 0.914 at $\alpha_{\text{test}} = 2$. The improvement was most pronounced at the extremes: at $\alpha_{\text{test}} = 0.125$, the PCC increased from 0.346 for the Phase Specialist to 0.772 and 0.858 for Phase Generalists I and II, respectively; at $\alpha_{\text{test}} = 2$, it increased from 0.615 to 0.746 and 0.914, respectively. As highlighted by the enlarged region around $\alpha_{\text{test}} = 1$ in **Fig. 9(b)**, this broader performance involved only a modest trade-off at the original operating point, where the PCC decreased from 0.974 for the Phase Specialist to 0.968 and 0.961 for Phase Generalists I and II, respectively.

These results reveal complementary phase-generalization capabilities in the optical encoder and digital decoder. Because the optimized optical encoder was jointly optimized only at $\alpha = 1$, its ability to support reconstruction across the complete tested interval, $\alpha_{\text{test}} \in [0.125, 2]$, shows that its compressed measurements preserve phase information beyond the range encountered during joint optimization/training. Decoder-only fine-tuning, together with expansion of the decoder output range, further improves structural fidelity of phase reconstructions over a broader range of object-phase amplitudes while leaving the optical encoder unchanged. The same physical encoder $\mathcal{E}_0$ can therefore be retained while the digital decoder is adapted to the phase range expected for a particular QPI application, avoiding the need to re-optimize or modify the optical system.

We next experimentally demonstrated the reuse of the same 20-pattern optical encoder across five phase-object datasets, including EMNIST and BloodMNIST, whose phase images were excluded from simulation-stage decoder training (**Figs. 10 and 11**). With the optical encoder $\mathcal{E}_0$ frozen, one multi-dataset decoder $\mathcal{D}_0$ was trained in simulation using the combination of MNIST, Fashion-MNIST, and OrganAMNIST, which we refer to as internal datasets because they were seen during decoder training. We then introduced EMNIST and BloodMNIST as external datasets that were excluded from this training stage, allowing us to test generalization to previously unseen object/data distributions. For this study, the simulation-stage decoder was trained using perturbations calibrated from experimental residuals computed on the experimental fine-tuning training sets. The held-out experimental test objects were excluded from residual calibration. The same decoder $\mathcal{D}_0$ was first applied directly to experimental measurements from all five datasets to establish zero-shot experimental results and was subsequently fine-tuned to produce $\mathcal{D}_{exp}$ using experimental data while keeping the optical encoder unchanged. This design therefore evaluates both simulation-to-experiment generalization for familiar object distributions and transfer learning to previously unseen data distributions using only digital decoder updates.

**Fig. 11** summarizes the dataset-transfer performance of the frozen optical encoder (values given as PCC and PSNR). For the internal datasets, **Fig. 11 (a,b)**, the simulation-trained decoder $\mathcal{D}_0$ transfers well to experimental measurements for MNIST and Fashion-MNIST test objects, with only a modest reduction in fidelity under zero-shot deployment: 0.895 (18.5 dB) and 0.918 (18.9 dB), compared with 0.945 (21.4 dB) and 0.942 (20.7 dB) in simulation, respectively. OrganAMNIST presents a more challenging data distribution: its simulated test performance is lower (0.728 (17.0 dB)), reflecting the more complex, low-contrast structure of these images, and its zero-shot performance is 0.679 (16.1 dB). Experimental decoder adaptation improves performance across all three datasets, with most of the gain occurring with $T$=50–100 transfer-learning objects per dataset before gradually approaching a plateau. At $T$ = 450, MNIST and Fashion-MNIST reach 0.942 (21.5 dB) and 0.937 (20.8 dB), matching their simulated performance, an increase of about 2–3 dB in PSNR over zero-shot. OrganAMNIST reaches 0.761 (17.5 dB), slightly exceeding its simulated value. The corresponding reconstructions show the same trend as in **Fig. 4**, with improved structural phase fidelity after adaptation while the optical encoder $\mathcal{E}_0$ remains unchanged.

These external phase datasets provide a more stringent test for our single-pixel QPI system because neither EMNIST nor BloodMNIST images were included in simulation-stage decoder training (Fig. 11**(c,d)**). Despite this distribution shift at the input phase objects, the zero-shot reconstructions retain recognizable object structure, 0.779 (15.1 dB) for EMNIST and 0.651 (13.2 dB) for BloodMNIST test images, indicating that the fixed encoder $\mathcal{E}_0$ preserves phase information beyond the object/data distributions used for the original encoder–decoder optimization. Subsequent experimental transfer learning substantially improves the QPI results of both datasets, with the largest gain observed for BloodMNIST, which rises from 0.651 (13.2 dB) to 0.884 (22.0 dB), reflecting an 8.8 dB improvement. Similarly, EMNIST improves from 0.779 (15.1 dB) to 0.871 (17.9 dB). In both cases, the adapted decoder $\mathcal{D}_{exp}(\boldsymbol{m}_{\text{exp}})$ surpasses the simulation-stage decoder's performance on the same test objects in simulation because the simulation-stage decoder was never trained on these distributions.

Together, the numerical benchmarks and optical experiments establish two complementary capabilities of learned optical compression for single-pixel QPI: improved phase-reconstruction fidelity over the tested prescribed encoders at the same measurement count, and reuse of the optimized optical encoder across new data distributions absent from its original training. This combination makes the learned measurement representation both compressive and reusable, rather than restricting it to the object distributions used for its initial optimization. Adapting the digital decoder not only compensates for changes in object phase statistics but also for simulation-to-experiment domain shifts. With the optical front-end held fixed, the system can therefore be extended to new phase-object data distributions through decoder fine-tuning alone, demonstrating the transferability and reusability of the optimized single-pixel QPI measurement framework.

# Discussion

In this work, we experimentally demonstrated single-pixel QPI with a jointly optimized optical encoder and digital decoder, reconstructing 28 × 28-pixel phase images of various test objects from 20 scalar intensity measurements. Under the same measurement budget, benchmarking showed that the optimized encoder substantially outperformed random, Hadamard, and Fourier patterns. With the same encoder states fixed, decoder-only adaptation improved experimental reconstruction fidelity and enabled transfer to previously unseen object phase distributions. Simulations further demonstrated that decoder retraining could improve structural reconstruction across a broader range of object-phase amplitudes. Together, these results establish a measurement-efficient and reusable optical encoding framework that can be adapted to different imaging conditions and object distributions without modifying the optical hardware.

The optical encoder of this single-pixel QPI architecture determines which information is retained in the scalar measurements and therefore places practical limits on reconstruction fidelity. By holding the decoder architecture and training protocol constant, the benchmark comparisons in **Fig. 7** empirically examine how the choice of optical encoding affects reconstruction performance under matched conditions. Each encoder was paired with an identical decoder architecture trained under identical conditions, so differences in QPI fidelity reflect the information supplied by each encoder type. Under these conditions, the optimized encoder reached a mean PCC of 0.969, compared with 0.811 for random, 0.804 for Hadamard, and 0.789 for Fourier patterns. The lower performance of the structured Hadamard and Fourier bases compared with random patterns indicates that structure alone does not guarantee effective encoding under the nonlinear phase-measurement model used here. Although Hadamard and Fourier bases are efficient for linear intensity imaging, this does not guarantee efficient phase encoding under a limited measurement budget. The optimized encoder instead allocates its 20 measurements according to both the object statistics and the measurement physics, as reflected in the spatial-frequency content of the optimized patterns (**Fig. 2(b)**). Optimizing the physical measurement is therefore not a refinement of the reconstruction but reflects a step that determines how much phase information the reconstruction can use. This distinction separates the present approach from earlier single-pixel QPI methods, which formed their measurements from prescribed rather than optimized patterns[36–38]. Such discrete-pattern implementations operated at measurement ratios of ~400–600%[37,38], whereas the present system performs quantitative phase reconstructions using a measurement ratio of $\rho = 2.55\%$.

Whereas the optical encoder determines which phase information is retained, the digital decoder determines how well that information is mapped to a phase reconstruction under experimental conditions. With the encoder fixed, decoder-only transfer learning using $T = 900$ objects per dataset substantially reduced the simulation-to-experiment mismatch. The same adaptation also shows what the encoder retains. The improvement in BloodMNIST reconstruction demonstrates that the frozen 20-measurement representation remains useful for phase-object distributions outside the encoder's original optimization set, with substantially improved reconstruction obtained through decoder-only adaptation. The present system therefore handles both experimental calibration and task changes by optimizing $\mathcal{D}_{exp}(\boldsymbol{m}_{\text{exp}})$, without re-optimizing the optical encoder $\mathcal{E}_0$. Because this flexibility rests entirely on the neural decoder, its size and computational cost determine how practical the approach is. The decoder maps the 20-element measurement vector to a reconstruction using 115.4 million trainable parameters and approximately 519.6 million multiply–accumulate operations per reconstruction. This is equivalent to approximately 1.04 GFLOPs when multiplication and addition are counted separately (see **Methods**). Although the decoder has many parameters, its computational demand per reconstruction is moderate: deep-learning QPI networks that recover phase from full-resolution intensity images or holograms typically require between about 10 and several hundred GFLOPs per image[53], because they process the full detector frame. Most of the decoder's parameters reside in a single fully connected layer (approximately 103 of 115 million), which makes it well suited to compression: low-rank factorization of large fully connected layers[54], weight pruning[55], or more parameter-efficient convolutional architectures[56] could shrink it substantially without changing the optimized optical measurement process.

In this time-multiplexed single-pixel QPI architecture, the number of measurements, $M$, controls the trade-off among compression, acquisition length, and reconstruction fidelity. As shown in **Fig. S1** (each $M$ trained separately with a reduced training budget; see **Supplementary Information**), increasing $M$ produces the largest gains in the strongly undersampled regime with low $\rho$ values. From $M = 2$ to $M = 10$, the combined PCC increases from ~0.291 to 0.869, while the corresponding PSNR rises from ~10 dB to 20 dB. Further increasing $M$ to 20 improves the combined PCC to ~0.942 and the PSNR to ~24 dB, whereas increasing the measurement count to $M = 50$ yields only modest additional gains. The dependence on $M$ is particularly pronounced for grating test images: their PCC is near zero at $M = 2$, compared with ~0.50 for MNIST, but increases to above 0.9 by $M = 20$, indicating that the grating-like test phase structures require greater measurement diversity under severe undersampling. These results therefore identify $M = 20$ as a practical operating point for the present system, capturing most of the achievable reconstruction improvement while limiting the sequential acquisition to 20 scalar measurements. For a $28 \times 28$-pixel phase reconstruction, this corresponds to a measurement ratio of $\rho = 0.0255$, a ~40× dimensional reduction.

Single-pixel detection at other parts of the electromagnetic spectrum could extend the significance of this approach beyond the visible-wavelength implementation demonstrated here. Although high-performance detector arrays are widely available at visible wavelengths, arrays can become substantially more expensive, less sensitive or more difficult to fabricate in infrared, terahertz and other spectral regions. Single-pixel architectures allow the use of detector technologies optimized for the wavelength, sensitivity, dynamic range or temporal bandwidth of interest without requiring those detectors to be fabricated as dense arrays[25,57]. They can also reduce detector readout and data-transfer requirements by recording a compact sequence of scalar values rather than spatially resolved intensity frames.

Future implementations could further improve the throughput and applicability of the presented single-pixel QPI framework. The sequential acquisition time could be reduced by using faster phase light modulators (PLMs)[58] and tighter synchronization between modulation and detection, since the present acquisition rate is primarily limited by the SLM update and settling time rather than by the optical measurement itself. Throughput could also be increased through spatial parallelization, by dividing the modulator aperture into independently optimized channels associated with separate single-pixel detector regions or single-element detectors, enabling parallel acquisition from multiple fields of view or phase objects. Beyond acquisition speed, wavelength diversity could increase the information available for reconstruction: because the phase delay introduced by a specimen depends on wavelength and material dispersion, measurements at multiple wavelengths provide additional constraints on the object, and diffractive multispectral QPI has previously produced wavelength-resolved quantitative phase maps across multiple bands[41]. A wavelength-diverse extension of the present system could therefore improve the robustness of the recovered phase or enable reconstruction of multispectral phase distributions. Finally, extending the current programmable-SLM validation to independent physical-phase specimens would provide a broader realization of the proposed single-pixel QPI architecture across various phase-imaging and sensing applications.

In summary, these results establish a controlled experimental foundation for measurement-efficient single-pixel QPI using a reusable optimized optical encoder. By jointly optimizing the physical measurement process and digital reconstruction, the proposed framework preserves quantitative phase information within a highly compressed set of scalar measurements while retaining adaptability through decoder-only retraining. More broadly, this work highlights the value of task-specific optical measurement design for quantitative phase imaging, particularly in regimes where spatially resolved detector arrays are costly, slow, or unavailable, or where reducing measurement and data requirements is itself advantageous.

# Methods

## Optical forward model

The single-pixel quantitative phase imaging system was represented by a differentiable optical encoder, $\mathcal{E}(x, y, t_i)$, followed by a digital decoder, $\mathcal{D}(\boldsymbol{m})$. The encoder $\mathcal{E}(x, y, t_i)$ used $M$ sequential phase masks, yielding one scalar measurement per mask, and was time-multiplexed over the acquisition times $t_1, \dots, t_M$, while the input phase object remained fixed. Let $\Phi_{\text{target}}(x, y)$ denote the target phase object to be reconstructed, and let $A(x, y)$ denote the uniformly illuminated aperture. At time $t_i$, the field at the input plane was

$$U_0(x, y, t_i) = A(x, y)\exp\big(j\big[\Phi_{\text{target}}(x, y) + \mathcal{E}(x, y, t_i)\big]\big). \tag{5}$$

Propagation through a homogeneous medium of refractive index $n_0$ over an axial distance z was written as the first Rayleigh–Sommerfeld diffraction integral,

$$U_z(x, y, t_i) = \iint U_0(x', y', t_i)h(x - x', y - y', z)dx'dy' \tag{6}$$

with

$$h(\Delta x, \Delta y, z) = \frac{z}{r^2}\left(\frac{1}{2\pi r} + \frac{1}{j\lambda_m}\right)\exp\left(\frac{j2\pi r}{\lambda_m}\right), \qquad r = \sqrt{\Delta x^2 + \Delta y^2 + z^2} \tag{7}$$

where $\lambda_m = \lambda/n_0$ is the wavelength in the propagation medium, with $\lambda$ denoting the vacuum wavelength. The numerical model evaluated the propagating-wave component using the angular-spectrum representation:

$$U_z = \mathcal{F}^{-1}\{\mathcal{F}\{U_0\}H_z\}, \tag{8}$$

where $\mathcal{F}\,\cdot$ and $\mathcal{F}^{-1}\,\cdot$ represent the Fourier and inverse Fourier transforms and $H_z(f_x, f_y)$ is

defined by

$$H_z(f_x, f_y) = \begin{cases} \exp(j2\pi z\sqrt{\left(\lambda_m^{(-2)} - f_x^2 - f_y^2\right)}), & f_x^2 + f_y^2 \le \lambda_m^{-2} \\ 0, & f_x^2 + f_y^2 > \lambda_m^{-2} \end{cases} \tag{9}$$

Average intensity detection at the detector plane over the single-pixel aperture $\Omega_D$ yields a single scalar measurement $m_i$:

$$m_i = \frac{1}{|\Omega_D|}\iint_{\Omega_D} |\,U_z(x, y, t_i)|^2\, dx\, dy \tag{10}$$

The complete acquisition yielded

$$\boldsymbol{m} = [\mathrm{m}_1, \dots, \mathrm{m}_\mathrm{M}]^\mathrm{T} \in \mathbb{R}^\mathrm{M} \tag{11}$$

For all the single-pixel QPI models the target phase map $\Phi_{\mathrm{target}}$ and reconstruction $\Phi_\mathrm{r}$ were defined on an $\mathrm{N} \times \mathrm{N}$ grid with $\mathrm{N} = 28$. The target was bilinearly interpolated onto an $\mathrm{N_o} \times \mathrm{N_o}$ object-plane grid with $\mathrm{N_o} = 100$ to be used as the input phase. The optical field was represented on an $\mathrm{N_s} \times \mathrm{N_s} = 2000 \times 2000$ simulation grid with pitch $\Delta_\mathrm{s} = 4\ \mu\mathrm{m}$. The encoder contained $\mathrm{N_e} \times \mathrm{N_e} = 500 \times 500$ independently controlled phase pixels with pitch $\Delta_\mathrm{e} = 8\ \mu\mathrm{m}$, and the interpolated object occupied $\mathrm{N_o} \times \mathrm{N_o} = 100 \times 100$ samples with pitch $\Delta_\mathrm{o} = 8\ \mu\mathrm{m}$. The numerical window therefore spanned $8.0 \times 8.0\ \mathrm{mm}^2$, the encoder region spanned $4.0 \times 4.0\ \mathrm{mm}^2$ and the interpolated object occupied $800 \times 800\ \mu\mathrm{m}^2$. The propagated light has a wavelength of $\lambda = 635$ nm, and $\mathrm{z}_0 = 5.2$ cm was used as the distance between the time-multiplexed encoder and the detector. The simulated detector has a length $w_d = 0.1\ mm$, or 25 simulation-grid samples.

### Modeling and initialization of optical time-multiplexed encoder

The phase-only encoder was parameterized by an unconstrained latent trainable tensor $\Theta(\mathrm{x}, \mathrm{y}, \mathrm{t_i})$, calculated by:

$$\mathcal{E}(\mathrm{x}, \mathrm{y}, \mathrm{t_i}) = 2\pi\sigma[\Theta(\mathrm{x}, \mathrm{y}, \mathrm{t_i})] \tag{12}$$

where $\sigma(\cdot)$ is the sigmoid function. The trainable variables were initialized as

$$\Theta(\mathrm{x}, \mathrm{y}, \mathrm{t_i}) \sim \mathcal{N}(0, \sigma_\Theta^2) \tag{13}$$

where $\mathcal{N}(\cdot)$ denotes a normal (Gaussian) distribution and $\sigma_\Theta$ is the standard deviation used to initialize the latent encoder parameters $\Theta$. The deployed model used $M = 20$ encoder states, $\sigma_\Theta = 0.5$ and $N_{\mathrm{SLM}} \times N_{\mathrm{SLM}} = 500 \times 500$ phase values, resulting in $MN_{\mathrm{SLM}}^2 = 5.0$ million optimized optical parameters. Once deployed, the same optimized encoder states $\mathcal{E}(\mathrm{x}, \mathrm{y}, \mathrm{t_i})$ were used for all reported experimental captures and were frozen during decoder transfer learning (**Fig. 2**).

### Evaluation metrics and statistical analysis

Reconstruction fidelity was quantified primarily using PCC and PSNR. Each reconstructed phase map and its ground-truth phase map were flattened before evaluation. For one pair, PCC was computed as

$$\mathrm{PCC}(\Phi_\mathrm{r}, \Phi_{\mathrm{target}}) = \frac{\sum_\mathrm{n} (\Phi_{\mathrm{r,n}} - \mu_\mathrm{r})(\Phi_{\mathrm{target,n}} - \mu_{\mathrm{target}})}{\sqrt{\sum_\mathrm{n} (\Phi_{\mathrm{r,n}} - \mu_\mathrm{r})^2 \sum_\mathrm{n} (\Phi_{\mathrm{target,n}} - \mu_{\mathrm{target}})^2}}$$

where n indexes pixels on the $28 \times 28$ reconstruction grid, and $\mu_\mathrm{r}$ and $\mu_{\mathrm{target}}$ are the mean values of the reconstructed $\Phi_\mathrm{r}$ and target $\Phi_{\mathrm{target}}$ phase maps, respectively. PCC was computed per object and then summarized across the corresponding blind-testing set.

MSE was computed between the reconstructed and target phase maps on the $N \times N$ reconstruction grid. For object $i$, the MSE was defined as

$$\mathrm{MSE_i} = \frac{1}{N^2}\sum_{x=1}^{N}\sum_{y=1}^{N}\left[\Phi_{r,i}(x,y) - \Phi_{target,i}(x,y)\right]^2$$

where $\Phi_{r,i}$ and $\Phi_{target,i}$ denote the reconstructed and target phase maps, respectively. For a test phase scale $\alpha_{\mathrm{test}}$, the target phase range was 0 to $\alpha_{\mathrm{test}}\pi$, and the peak value used to calculate phase PSNR was $L = \alpha_{\mathrm{test}}\pi$ radians:

$$\mathrm{PSNR} = 10\log_{10}\left(\frac{(\alpha_{\mathrm{test}}\pi)^2}{\mathrm{MSE}}\right)$$

For grating analyses, phase-error profiles were computed along selected one-dimensional sections through the reconstructed and target phase maps.

## Numerical benchmarking of optical encoding strategies

Four phase encoding strategies were compared numerically: optimized, random, Hadamard, and Fourier phase encoding. Every encoder contained $M = 20$ phase-only masks, each represented by 500 × 500 independently specified SLM phase values, and produced one scalar intensity measurement per mask. The optimized condition used the original encoder–decoder pair, $\mathcal{E}_0 - \mathcal{D}_0$, obtained through end-to-end joint optimization and subsequently deployed in the experimental system. For each fixed-pattern baseline, the random, Hadamard, or Fourier masks were frozen and paired with a newly initialized digital decoder having the same architecture as $\mathcal{D}_0$. Each baseline decoder was then trained from scratch to reconstruct the phase objects from the measurements produced by its corresponding encoder. The comparison therefore evaluates the performance of the jointly optimized optical–digital system against conventional fixed optical encoders followed by independently optimized digital reconstruction.

The random encoder consisted of independently sampled phase values drawn uniformly from $[0, 2\pi]$. The Hadamard masks were generated from a $512 \times 512$ Sylvester Hadamard matrix[63], which was the smallest power-of-two construction covering the $500 \times 500$-pixel SLM grid. Each two-dimensional pattern was formed as the outer product of two Hadamard rows, cropped to $500 \times 500$ pixels, and mapped from binary values to phase levels of 0 and π. Two Hadamard orderings were evaluated: an index-based Sylvester ordering and a cake-cutting ordering based on the number of connected regions in each pattern. The Sylvester ordering produced the higher reconstruction PCC and was used for the comparison reported in **Figs. 6** and **7**. Fourier masks were constructed from two-dimensional cosine patterns,

$$\phi_{u,v,\delta}(x,y) = \frac{\pi}{2}\left[1 + \cos\left(2\pi\frac{ux + vy}{S} + \delta\right)\right]$$

where $S = 500$, and $(u, v)$ denotes the spatial-frequency index. Frequencies were selected from a $(6 \times 6)$ grid, ordered by increasing $u^2 + v^2$, with the zero-frequency component omitted. We evaluated a cosine-only implementation, which assigned one mask to each of the 20 lowest nonzero frequencies, and a quadrature implementation, which assigned phase offsets of $\delta = 0$ and $\pi/2$ to each of the ten lowest frequencies. The quadrature implementation achieved the higher PCC and was used as the Fourier baseline reported in **Figs. 6-7**. Both the Hadamard ordering and the Fourier variant were selected based on reconstruction PCC on the validation set.

**Supplementary Information:**

Supplementary Information includes:

- **Supplementary Fig. S1**
- **Digital decoder architecture**
- **Loss functions for single-pixel QPI joint optimization**
- **Experimental implementation details**
- **Experimental transfer learning**
- **Datasets and preprocessing**
- **Experimental noise calibration and noise-augmented simulations**

# Figures

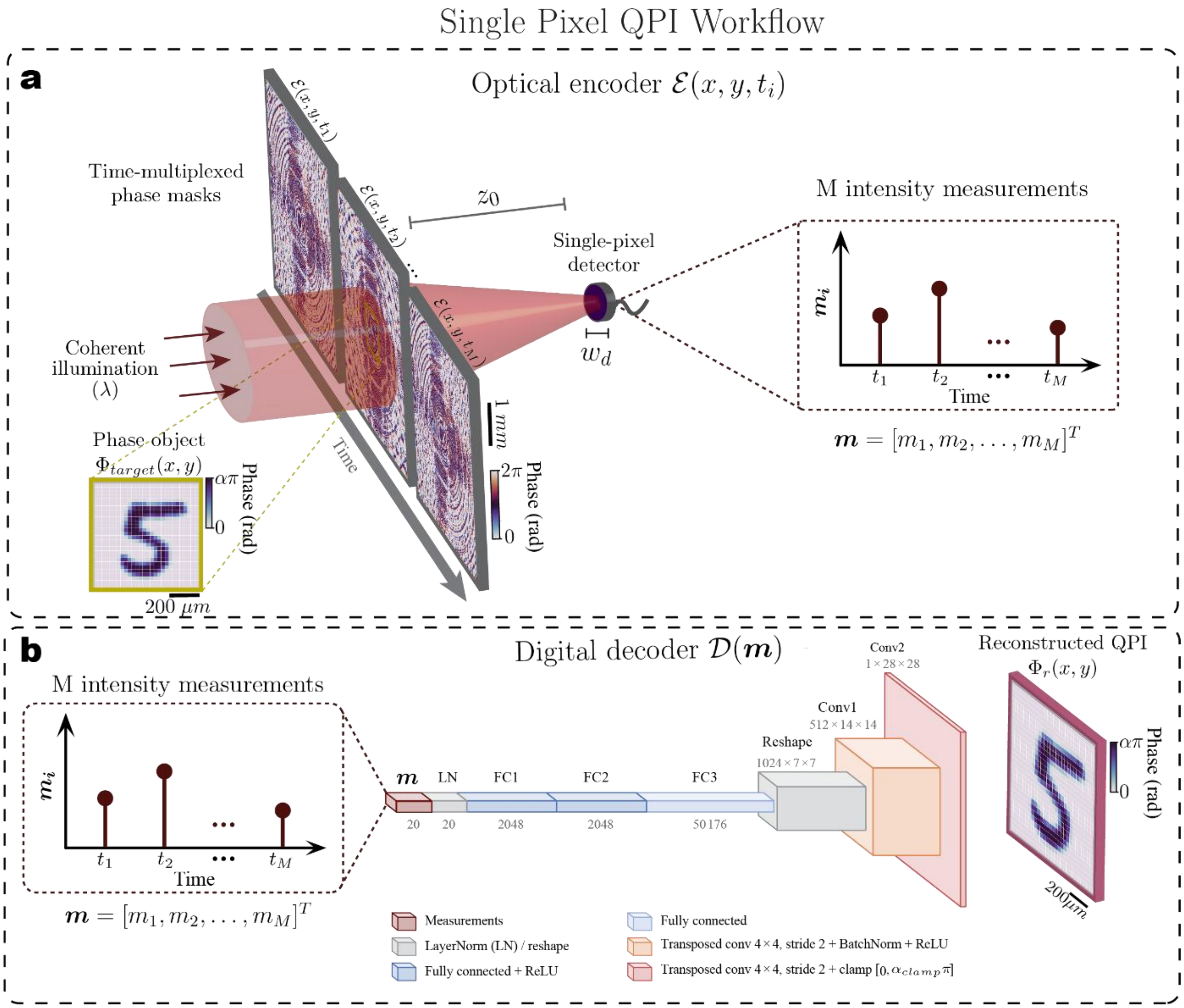

**Figure 1. Time-multiplexed workflow of learned optical compression for single-pixel QPI. (a)** A coherently illuminated phase object, $\Phi_{\text{target}}(x, y)$, is superimposed onto a sequence of $M$ optimized phase masks, $\mathcal{E}(x, y, t_i)$, displayed sequentially ($i = 1, \ldots, M$); these phase masks are frozen after optimization and remain the same, independent of the tested phase pattern to be reconstructed. For each encoder state, the propagated optical intensity is averaged over a single-pixel detector of width $w_d$, producing one intensity measurement $m_i$. These single-pixel measurements form the vector $\boldsymbol{m} = [m_1, m_2, \ldots, m_M]^T$. **(b)** The digital decoder, $\mathcal{D}(\mathbf{m})$, maps the measurement vector through normalization, fully connected layers and transposed-convolutional layers to reconstruct the quantitative phase distribution, $\Phi_r(x, y) \approx \Phi_{\text{target}}(x, y)$.

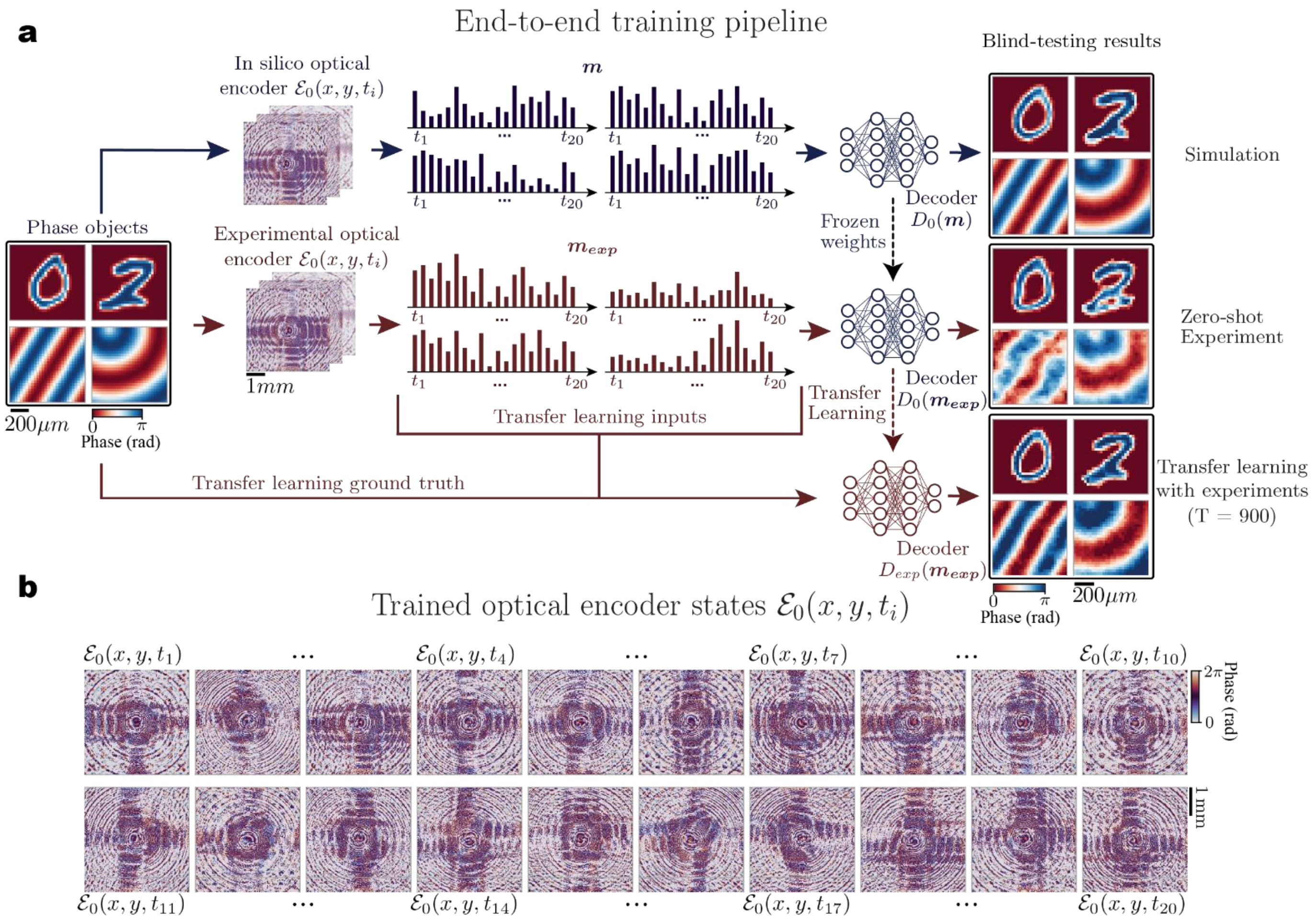

**Figure 2.** Schematic of end-to-end training and deployment of the optimized optical encoder. **(a)** Simulated phase objects are processed by a jointly optimized optical encoder $\mathcal{E}_0$ and a digital decoder $\mathcal{D}_0$. The same frozen encoder and decoder are then evaluated zero-shot using experimental measurements. Separate experimental measurement-target pairs, excluded from blind testing, are subsequently used to fine-tune the experimental decoder, i.e., $\mathcal{D}_0 \rightarrow \mathcal{D}_{exp}$, that accounts for system-specific measurement discrepancies. **(b)** Visualization of the $M = 20$ optimized encoder phase masks.

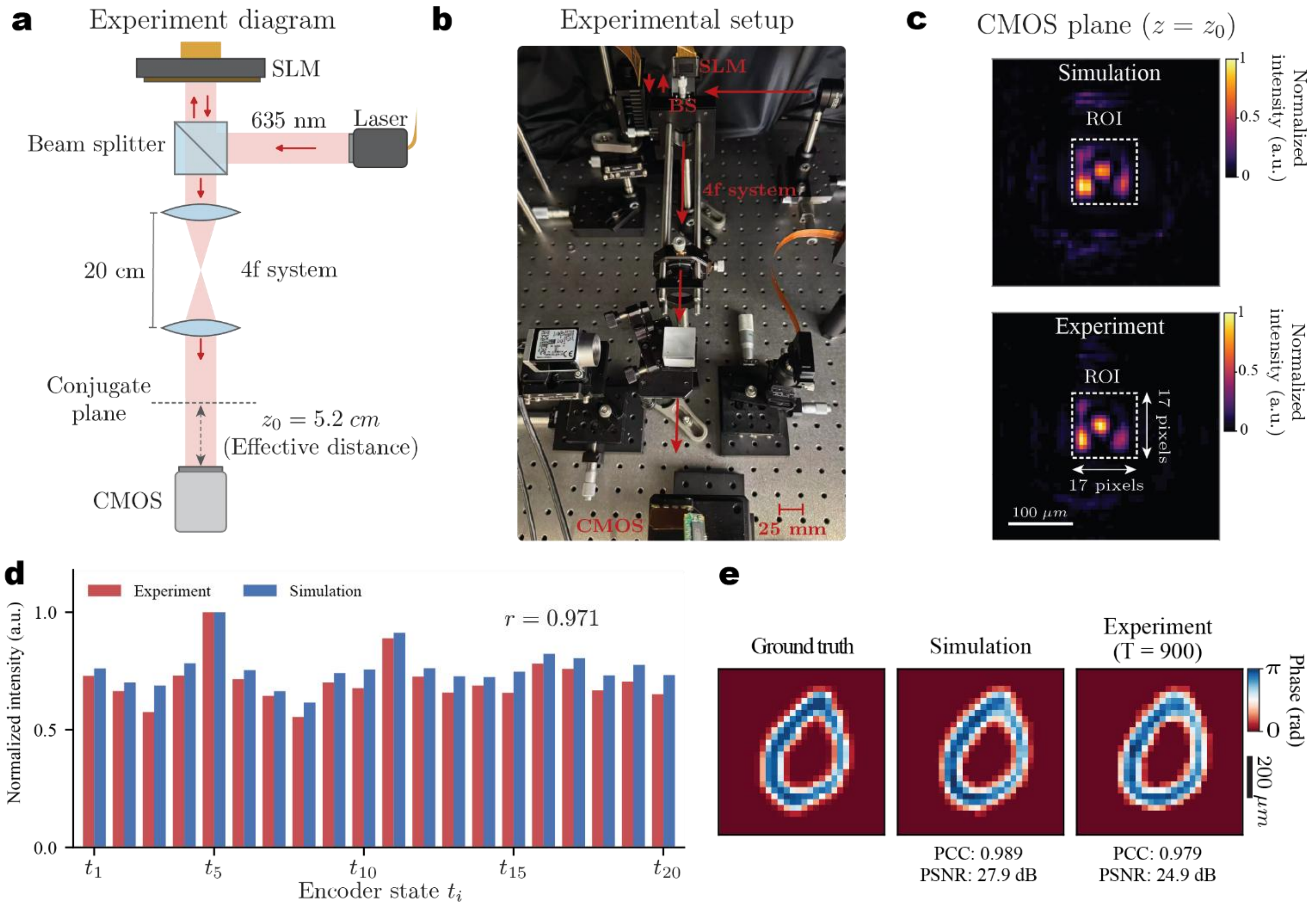

**Figure 3.** Experimental setup for single-pixel QPI. **(a)** Optical schematic. A 635 nm laser beam illuminates a phase-only SLM through a beam splitter; the reflected beam is relayed by a 4f system to a plane conjugate to the SLM, and the CMOS camera is placed at an effective distance $z_0 = 5.2\ cm$ beyond it. The camera emulates the single-pixel detector: each measurement is the intensity averaged over a region of interest (ROI) with $w_d = 0.1\ mm$. **(b)** Photograph of the experimental setup. **(c)** Detector-plane intensity for one encoder state ($t_1$), simulated (top) and measured with the CMOS camera (bottom), shown over the same 0.35 mm field centered on the optical axis and normalized to the maximum of each image; the dashed square marks the 100 µm detector region. **(d)** The $M = 20$ single-pixel measurements of the object in **(e)**, experiment and simulation, each normalized to its maximum (Pearson correlation r = 0.971). **(e)** Target phase object (an MNIST digit held out from training) and its reconstructions from simulated measurements, using the simulation-trained decoder, and from experimental measurements, after transfer learning with $T = 900$ experimental objects per dataset; PCC and PSNR are given below each reconstruction.

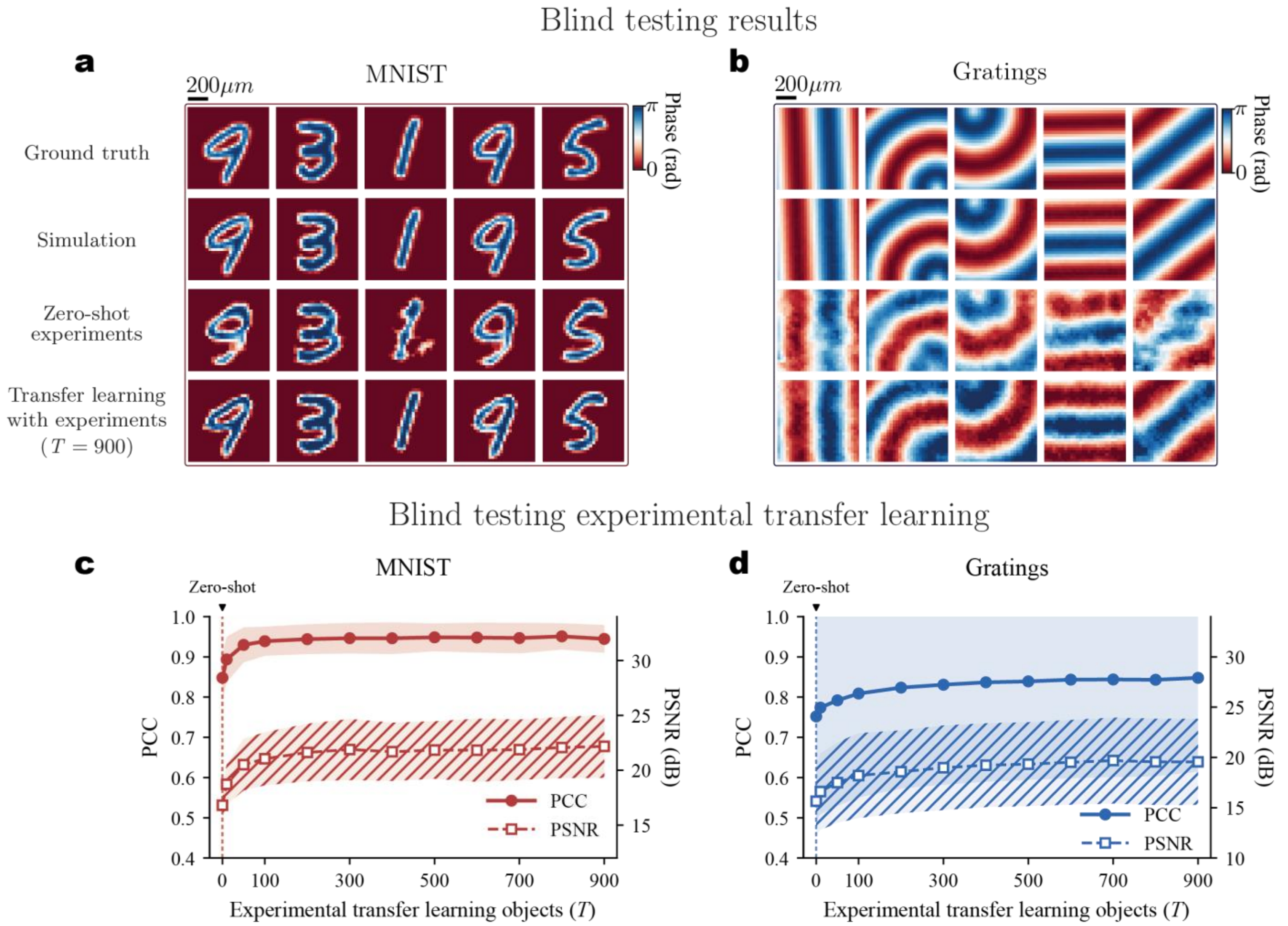

**Figure 4. Experimental transfer learning for MNIST and grating-like phase test objects. (a,b)** Blind-testing reconstructions of MNIST and grating-like phase test objects, respectively, comparing the ground truth, simulation reconstruction, zero-shot experimental reconstruction, and experimental reconstruction after transfer learning with $T = 900$ experimental objects per dataset (900 MNIST and 900 grating objects). **(c,d)** Mean Pearson correlation coefficient (PCC, left axis, filled circles) and peak signal-to-noise ratio (PSNR, right axis, open squares) as functions of the number ($T$) of experimental transfer-learning objects per dataset, for MNIST and gratings, respectively; $T = 0$ corresponds to zero-shot reconstruction. A single digital decoder was fine-tuned on balanced numbers of objects from both datasets while the optical encoder remained fixed. Lines show the mean over 100 blind-testing experimental objects per dataset, and shaded regions indicate ±1 standard deviation (hatched for PSNR). From $T = 0$ to $T = 900$, the mean PCC increased from 0.848 to 0.945 for MNIST and from 0.752 to 0.848 for gratings, and the mean PSNR increased from 16.8 dB to 22.2 dB and from 15.7 dB to 19.6 dB, respectively.

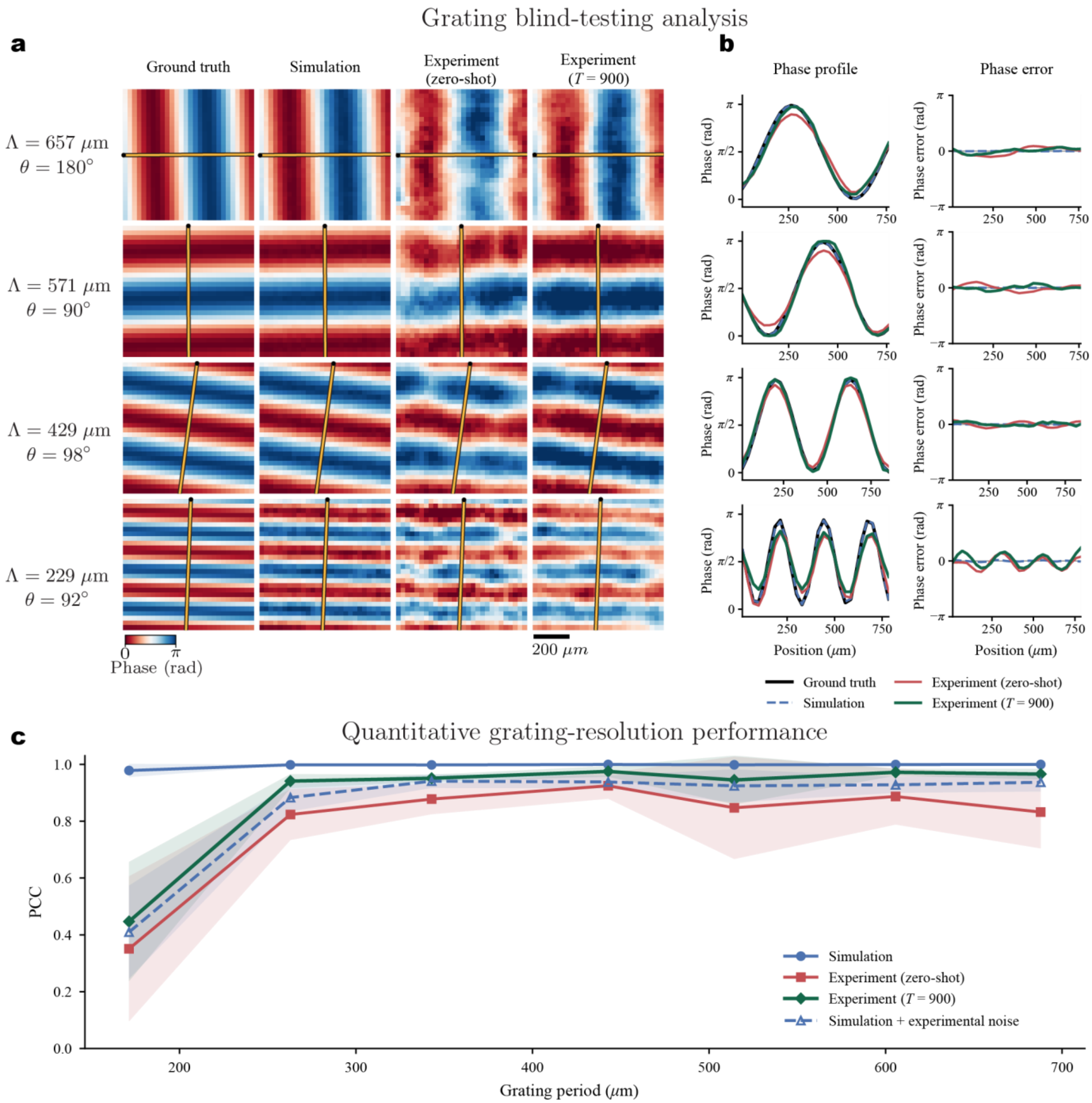

**Figure 5. Quantitative grating resolution and phase-error analysis. (a)** Representative linear gratings with periods $\Lambda = 657$, 571, 429 and 229 μm, shown as ground truth alongside the simulated reconstructions and two experimental reconstructions: zero-shot and after transfer learning with $T = 900$ experimental captures per dataset. All four gratings were held out from the transfer-learning set. Yellow lines mark where the phase profiles are extracted, oriented normal to the grating fringes. **(b)** Phase profiles extracted along the marked lines for the ground truth, the simulated reconstructions and both experimental reconstructions, together with the corresponding phase-error profiles relative to the ground truth. **(c)** Reconstruction PCC as a function of grating period for ideal simulation, simulation with calibrated experimental noise ($\sigma_m$= 0.313), and the zero-shot and transfer-learned ($T = 900$) experimental reconstructions. All curves are evaluated on the same 54 linear gratings held out from transfer learning. Points show the mean PCC within each of 7 grating-period bins, and shaded regions indicate ±1 standard deviation. The noise-calibrated simulation reproduces much of the degradation observed experimentally at shorter grating periods, and transfer learning improves the experimental reconstructions at every period.

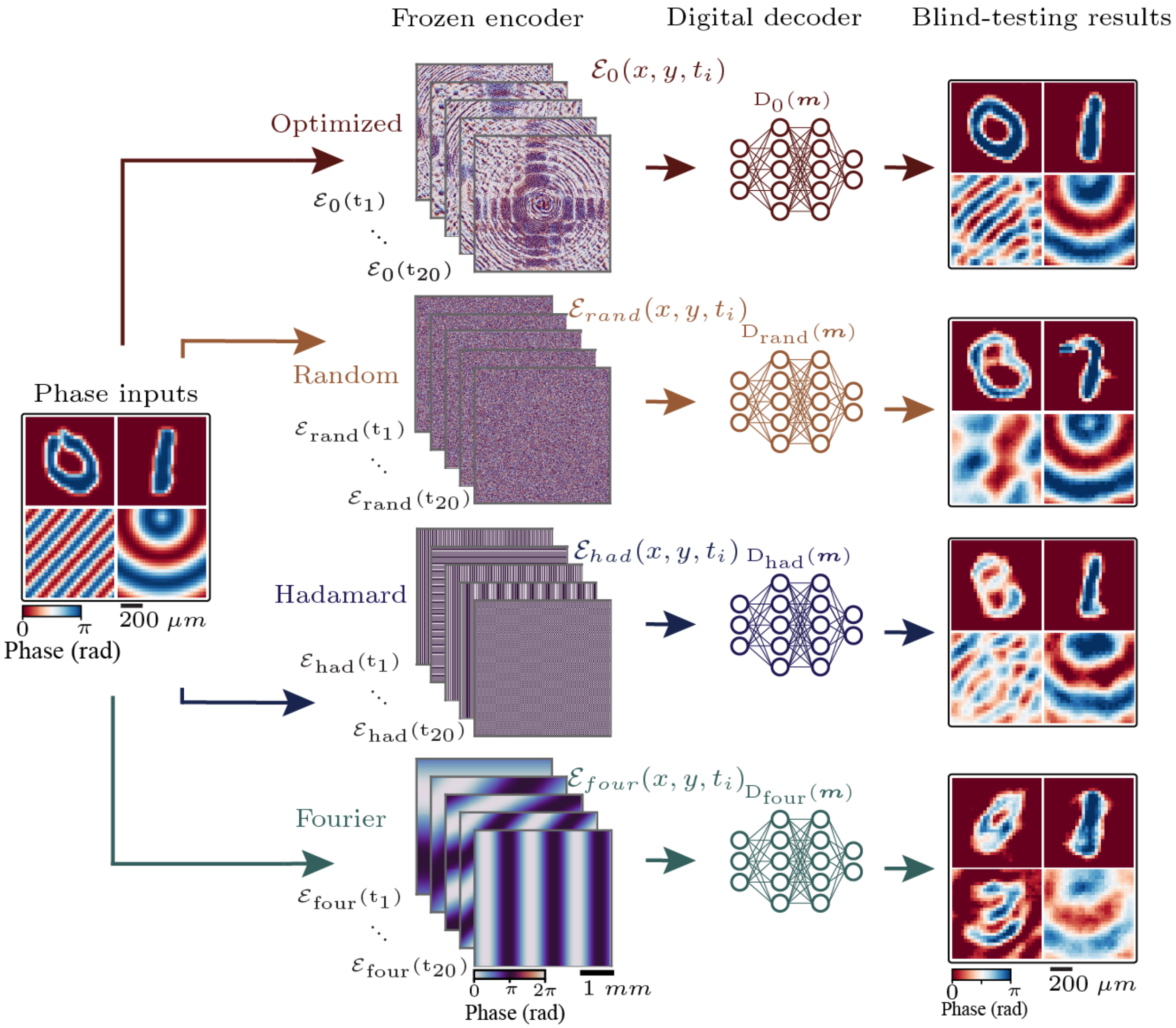

**Figure 6. Benchmark comparisons for optimized and fixed-pattern phase encoders.** Identical phase objects are encoded using $M = 20$ optimized ($\mathcal{E}_0$), random, Hadamard, or Fourier phase patterns. Each frozen encoder produces a corresponding scalar measurement vector $\boldsymbol{m}$, which is processed by its respective digital decoder. Reconstructions are evaluated on blind-testing phase objects not used during optimization.

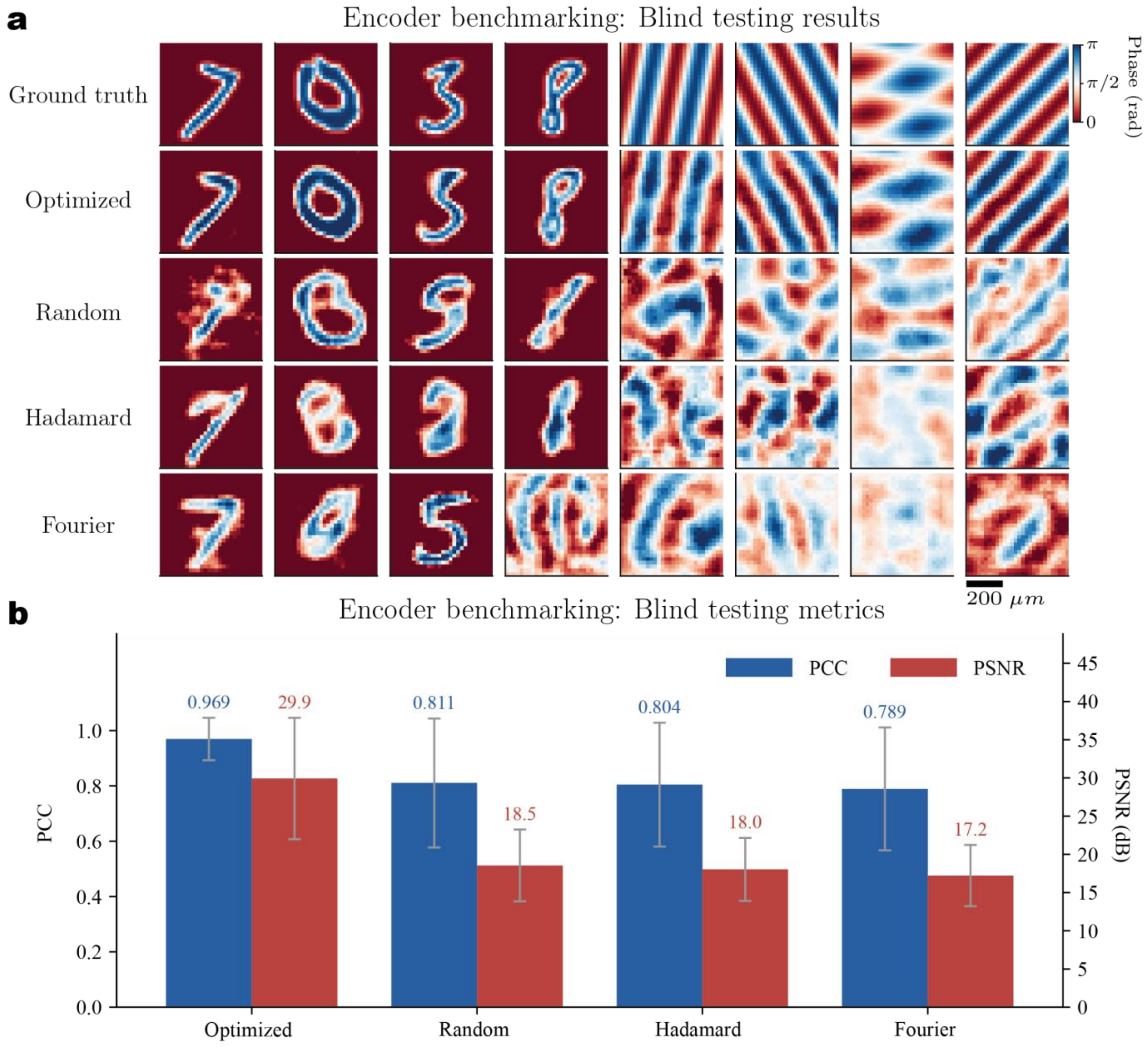

**Figure 7. Statistical comparison of the optimized optical encoder against fixed-pattern encoders. (a)** Target phase objects and blind-testing reconstructions obtained using optimized ($\mathcal{E}_0$), random, Hadamard, and Fourier phase encoders under the same $M = 20$ measurement budget. The optimized encoder was paired with its original jointly trained digital decoder, whereas each frozen fixed-pattern encoder was paired with a newly initialized digital decoder of identical architecture that was optimized independently under the same training protocol. The displayed examples include MNIST and grating phase test objects; the aggregate metrics include all 2000 blind-testing objects. **(b)** Mean PCC and PSNR values for each encoder over the 2000 blind-testing objects; error bars indicate ±1 standard deviation across objects. For the random encoder, reconstruction metrics were first averaged over five independently generated mask sets for each test object; the reported means and standard deviations were then calculated across all test objects.

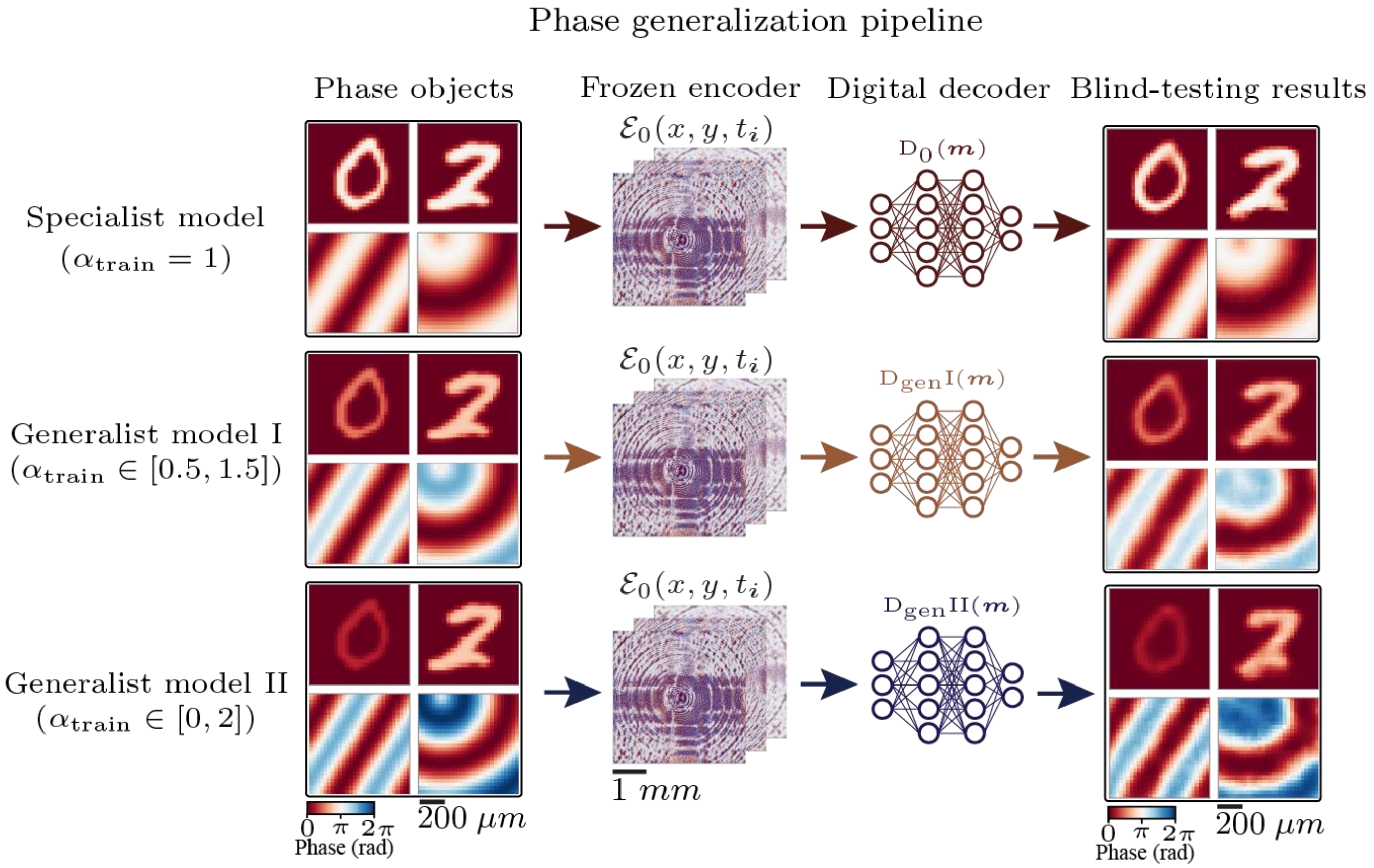

**Figure 8. Optimization of specialist and generalist digital decoders for internal and external generalization in numerical simulations**. The same frozen optimized optical encoder ($\mathcal{E}_0$) is used to produce $M = 20$ scalar measurements for all phase-range conditions. The specialist digital decoder $D_0$ is optimized using phase objects with a fixed scaling factor of $\alpha_{\text{train}} = 1$, corresponding to phase values spanning 0 to $\pi$. Generalist I is initialized from the specialist decoder and further optimized using phase objects with ($\alpha_{\text{train}} \in [0.5,1.5]$), while Generalist II is initialized from the same specialist decoder and optimized over the broader interval $\alpha_{\text{train}} \in [0,2]$. The phase-scaling factor is sampled from the corresponding interval during optimization, while the optical encoder ($\mathcal{E}_0$) remains unchanged. The three decoders are subsequently evaluated across maximum phase values from $0.125\pi$ to $2\pi$ ($\alpha_{\text{test}} \in [0.125, 2]$) to compare QPI reconstruction fidelity within their optimization ranges, representing internal generalization, and beyond those ranges, representing external phase-range generalization; also see **Fig. 9** for this comparison.

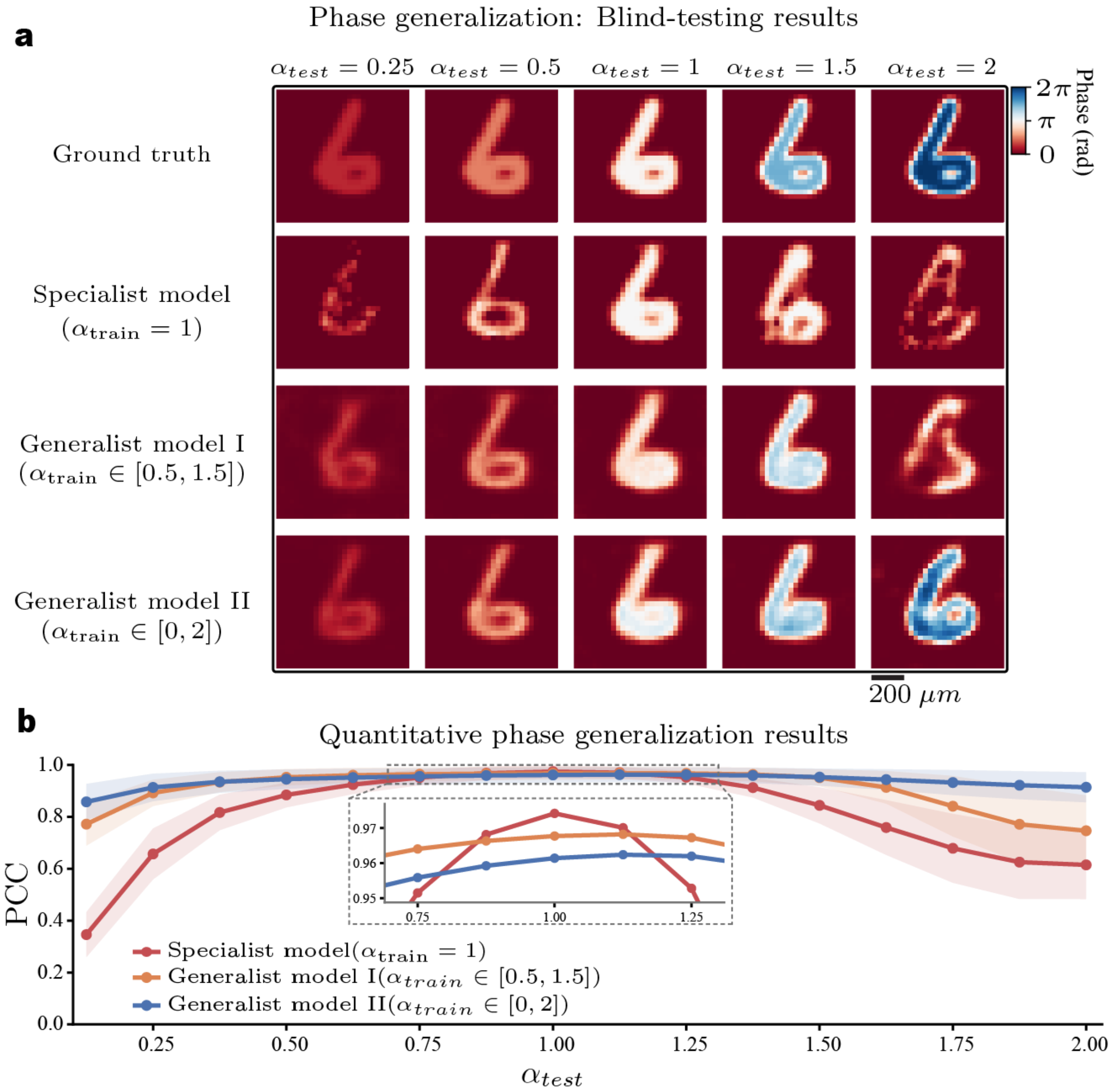

**Figure 9. Numerical evaluation of internal and external phase-range generalization of specialist and generalist digital decoders.** (a) Target phase objects and reconstructed MNIST objects for maximum phase values of $\alpha_{test}\pi$, with $\alpha_{test} = 0.25, 0.5, 1, 1.5$, and 2. Reconstructions are compared for a specialist decoder optimized at $\alpha_{train} = 1$, Generalist I decoder optimized with $\alpha_{train} \in [0.5,1.5]$, and Generalist II decoder optimized with $\alpha_{train} \in [0,2]$. All three digital decoders use the same frozen optimized optical encoder ($\mathcal{E}_0$). (b) Mean PCC as a function of $\alpha_{test}$ for the 10,000 blind-testing MNIST objects. Lines show the mean, shaded regions indicate ±1 standard deviation across objects, and the inset magnifies the region near $\alpha_{test} = 1$. Performance within the decoder training interval characterizes internal generalization, whereas performance beyond that interval characterizes external phase-range generalization.

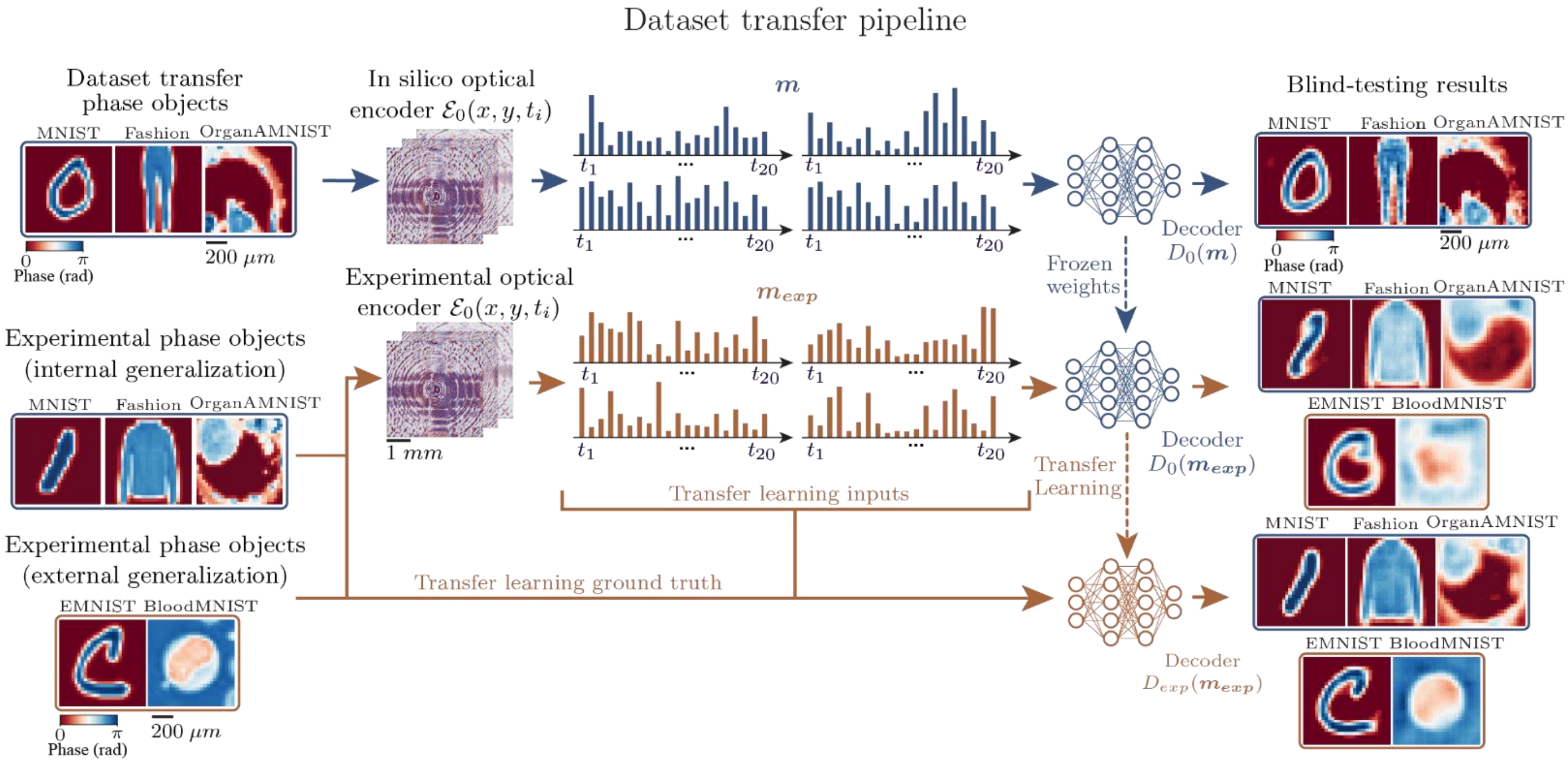

**Figure 10. Experimental transfer learning and external generalization with a frozen optimized optical encoder ($\mathcal{E}_0$).** Simulated phase objects drawn from MNIST, Fashion-MNIST, and OrganAMNIST are processed by the frozen optical encoder to generate measurement vectors $\boldsymbol{m}$, which were subsequently augmented with calibrated perturbations (see **Methods**). These measurements are used to optimize the simulation-stage digital decoder $\mathcal{D}_0$. The same optical encoder is then deployed experimentally, and the resulting experimental measurement vectors $\boldsymbol{m_{exp}}$ are first reconstructed zero-shot using $\mathcal{D}_0$. Experimental transfer learning is subsequently performed using measurements from MNIST, Fashion-MNIST, OrganAMNIST, EMNIST, and BloodMNIST while the optimized optical encoder remains frozen. The EMNIST and BloodMNIST image distributions were excluded from simulation-stage reconstruction training and were used to evaluate external generalization.

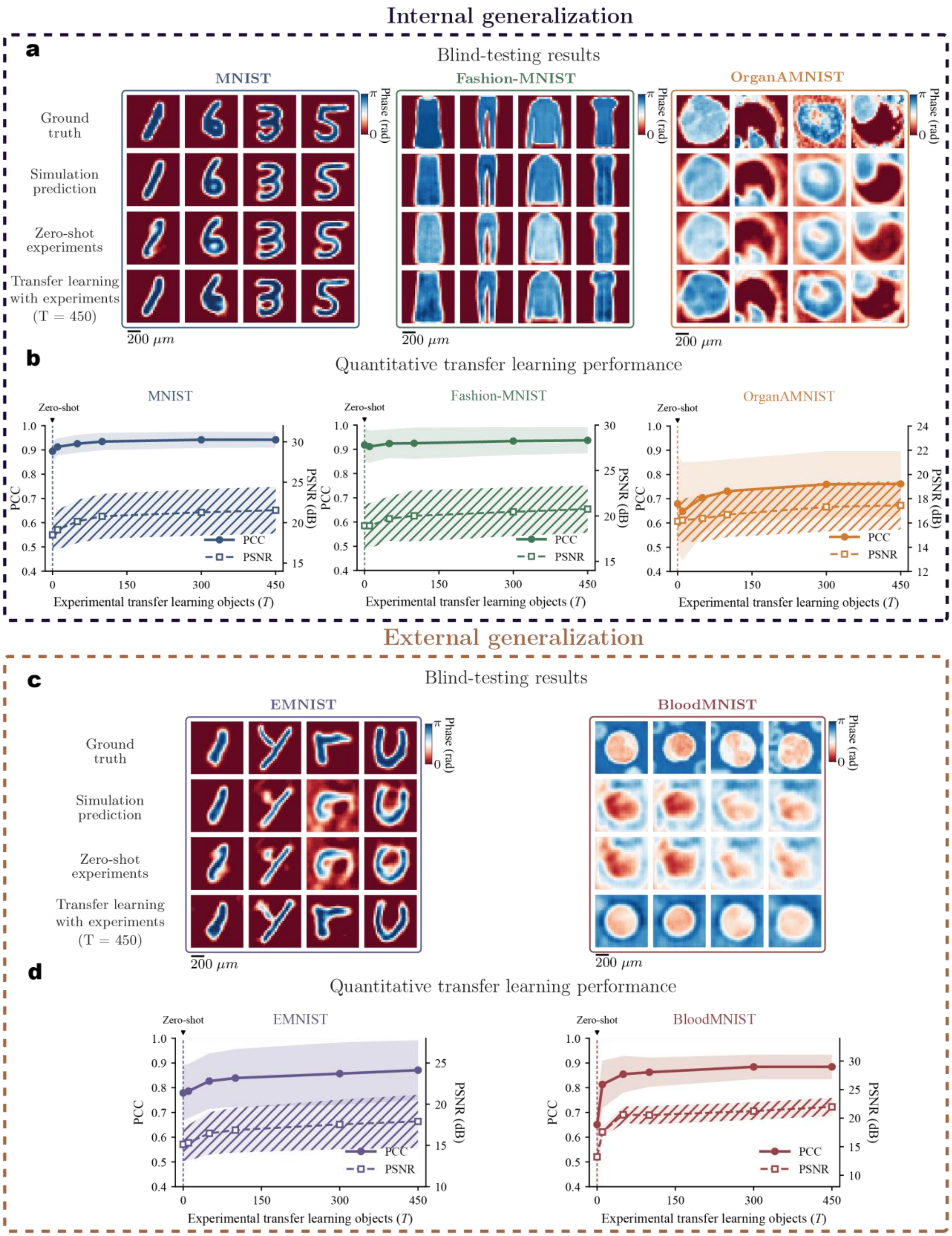

**Figure 11. Analysis of decoder transfer learning for internal and external datasets under a frozen optical encoder. (a)** Visualization of the reconstructed phase images of samples never seen

before, sampled from internal datasets MNIST, Fashion-MNIST, and OrganAMNIST, which are used during in silico training, comparing ground truth, simulation prediction, zero-shot experimental reconstruction, and experimental transfer-learning reconstructions with $T = 450$. **(b)** Mean PCC (left axis, filled circles) and PSNR (right axis, open squares) against the number of experimental transfer-learning objects per dataset, for the internal datasets; $T = 0$ corresponds to zero-shot reconstruction, which denotes experimental reconstruction without decoder fine-tuning on paired experimental measurements and phase targets. **(c)** Corresponding visualization results for the external datasets, i.e., EMNIST and BloodMNIST datasets, which were unseen during simulation training. **(d)** Mean PCC and PSNR against the number of experimental transfer-learning objects per dataset, for the external datasets. Curves summarize 50 blind-testing experimental captures per dataset. Shaded regions show ±1 standard deviation (hatched for PSNR). The encoder ($\mathcal{E}_0$) remained frozen throughout, and only decoder weights were adapted.